\documentclass[twocolumn,prl,superscriptaddress,showpacs]{revtex4}
\usepackage{graphicx,amsmath,amssymb,color}
\usepackage{dcolumn}
\usepackage{bm}
\usepackage[ansinew]{inputenc}

\begin{document}

\title {Universal phase relation between longitudinal and transverse fields observed in focused terahertz beams}

\author{S. Winnerl}
\email{s.winnerl@hzdr.de}
\affiliation{%
Helmholtz-Zentrum Dresden-Rossendorf, P.O. Box 510119, 01314 Dresden, Germany
}%
\author{R. Hubrich}
\affiliation{%
Helmholtz-Zentrum Dresden-Rossendorf, P.O. Box 510119, 01314 Dresden, Germany
}%
\author{M. Mittendorff}
\affiliation{%
Helmholtz-Zentrum Dresden-Rossendorf, P.O. Box 510119, 01314 Dresden, Germany
}%
\affiliation{%
Technische Universität Dresden, 01062, Dresden, Germany
}%
\author{H. Schneider}
\affiliation{%
Helmholtz-Zentrum Dresden-Rossendorf, P.O. Box 510119, 01314 Dresden, Germany
}%
\author{M. Helm}
\affiliation{%
Helmholtz-Zentrum Dresden-Rossendorf, P.O. Box 510119, 01314 Dresden, Germany
}%
\affiliation{%
Technische Universität Dresden, 01062, Dresden, Germany
}%

\date{\today}

\begin{abstract}
We directly observe longitudinal electromagnetic fields in focused freely propagating terahertz beams of radial and linear polarization. In accordance with theory, the longitudinal fields are phase shifted by a value of 
$\pi/2$ with respect to the transverse field components. This behavior is found for all frequency components of single cycle THz radiation pulses. Additionally we show that the longitudinal field of a radially polarized THz beam has a smaller spot size as compared to the transverse field of a linearly polarized beam, that is focused under the same conditions. 
\end{abstract}

\maketitle

%



In many textbooks light is treated as a plane wave, which has purely transverse character. In reality, except for azimuthally polarized beams, any beam of finite diameter exhibits longitudinal components \cite{Sheppard1999}. In the focus the strength of the longitudinal fields relative to the transverse fields is enhanced. Longitudinal components were first observed in focused linearly polarized microwave beams \cite{Carswell1965}. While in that experiment the intensity of the longitudinal components was about 400 times smaller than the intensity of the transverse components, focussing radially polarized beams of visible light with high numerical aperture yielded intensities of the longitudinal components exceeding the intensity of the transverse components \cite{Novotny2001, Dorn2003}. Such beams exhibit a smaller spot size as compared to Gaussion beams of linear polarization \cite{Quabis2000, Novotny2001}. The observation of such strong and tightly focused longitudinal electromagnetic waves stimulated experimental \cite{Oron2000, Moshe2003, Maurer2007, Chang2007, Grosjean2008, Winnerl2009, Tavalla2011, Salamin2008} and theoretical \cite{Hernandez2006, Li2007, Urbach2008, Wang2008} work on the generation of radially polarized beams and on their specific properties. Already in 1959 Richards and Wolf showed that the longitudinal field components of linearly polarized waves obey a phase relation of $\pi/2$ with respect to the transverse field components \cite{Richards1959}, later Youngworth and Brown found the same result for radially polarized beams \cite{Youngworth2000}. Experimentally this relation has not been verified yet, since so far in all experimental studies the intensity  has been measured. Hence those experiments can not provide phase information. In the terahertz (THz) spectral range field-resolved measurements revealing the phase information are a widely applied technique \cite{Han2001}. Furthermore the method of electro-optic sensing employed to register the THz fields is capable to distinguish between longitudinal and transverse field components. This has been demonstrated by detecting longitudinal fields in the near field of a metallic tip \cite{Valk2002}. In contrast to such THz fields associated with surface-plasmon polaritons, we investigate longitudinal THz fields of freely propagating waves of radial and linear polarization, and in particular we measure the phase relation between longitudinal and transverse field components. We observe the predicted value $\pi/2$ for both radially and linearly polarized pulsed THz radiation. A symmetry argument is provided to reveal the universal character of this phase relation. The experimental results are compared with theory describing the beam properties beyond the paraxial approximation. A smaller spot size of the longitudinal components of a radially polarized beam compared to a standard Gaussian beam is observed even for moderate focusing conditions.

\begin{figure}
\includegraphics[width=8cm]{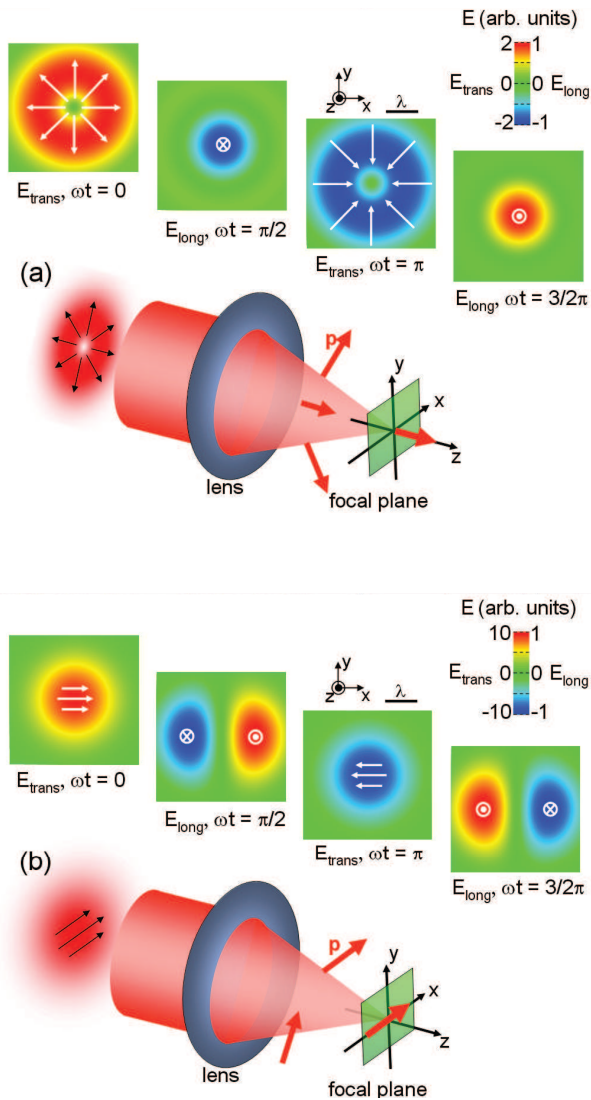}
\caption{\label{fig:epsart} (Color online) Schematic representation of focused beams of radial (a) and linear (b) polarization.  In the small panels the calculated field distribution in the focal plane is depicted for different phases $\omega t$. Note that the direction of the transverse components is encoded in the false color in a different way for the two modes, namely in radial direction for the radially polarized beam and along the $x$-axis for the linearly polarized beam. A scale bar indicates the wavelength $\lambda$.}
\end{figure}

Due to their symmetry radially polarized beams offer a particularly intuitive access to understanding longitudinal fields and their phase relation with respect to transverse fields. In Fig. 1a a radially polarized continuous wave, that is travelling along the $z$-direction and focused by a lens, is depicted. The unfocused radially polarized wave has a donut-like intensity distribution and belongs to the class of Bessel-Gauss beams, which are solutions of the vectorial Helmholtz equation \cite{Hall1996}. Already a simple geometric optics argument suggests that longitudinal fields should appear in the focus. As the wavefronts get tilted by the focussing element, the polarization vector $\mathbf{p}$ exhibits components along the $z$-direction that add up in the focus (cf. Fig. 1a). The four small panels depict calculated field patterns in the focal plane for different phases of the wave. For $\omega t = 0$ the field distribution is purely transverse with the field pointing radially outwards \cite{comment1}. A two-dimensional field pattern of this structure would require a charge in the center as a source of the field, hence it is forbidden by Maxwell's law $\textrm{div} \mathbf{E} = 0$ in vacuum. In vacuum, the field pattern has to extend in the $z$-direction and a longitudinal field on the axis is required to provide an influx into the center of the radial pattern of the transverse components. Indeed, for the phase  $\omega t = \pi/2$, the field is purely longitudinal over the whole focal plane with a pronounced maximum on the axis of propagation. The field is pointing in negative $z$-direction und thus provides the required flux. For the phase  $\omega t = \pi$ ($\omega t = 3/2\pi$) purely transverse (purely longitudinal) patterns occur with inverted signs of the fields. Due to the symmetry of the transverse beam for $\omega t = 0$ and $\omega t = \pi$ the maximum of the longitudinal field has to occur exactly for $\omega t = \pi/2$. We note that for phases that are not integer multiples of $\pi/2$ longitudinal and transverse fields occur simultaneously in the focal plane. Fig. 1b shows a linearly polarized continuous wave focused under the same conditions as for the radially polarized wave. Again purely transverse \cite{comment2} (longitudinal) fields occur in the focal plane for $\omega t = 0, \pi, 2\pi,... $ ($\omega t = 1/2\pi, 3/2\pi,... $). The appearance of longitudinal fields even for linearly polarized beams can be understood as a consequence of $\textrm{div} \mathbf{E} = 0$ for beams of finite diameter. For example the transverse field pointing in positive $x$-direction for $\omega t = 0$ requires on the right side of the pattern an outflux, which is provided by the longitudinal fields pointing in positive $z$-direction for $\omega t = \pi/2$ and in negative $z$-direction for $\omega t = -\pi/2$ (equivalent to $\omega t = 3/2\pi$). The relative strength of the longitudinal fields compared to the transverse fields is much smaller for the linearly polarized beam as compared to the radially polarized beam (cf. differnt false color scales for $E_{long}$ and $E_{trans}$ in Fig. 1a and 1b). While expressions in closed form for the longitudinal and transverse field components in the focal plane can be derived only of fundamental modes of certain symmetry, the $\pi/2$ phase difference between these components is universal, as any transverse pattern with a finite divergence in the focal plane will require longitudinal components. For symmetry reasons their phase has to be shifted by $\pi/2$. 

In our experiment we generate radially and linearly polarized single-cycle THz pulses by exciting photoconductive structures with near-infrared femtosecond laser pulses (wavelength 800 nm, repetition rate 78 MHz, duration 50 fs, pulse energy 8 nJ). A microstructured interdigitated electrode structure, consisting of linear stripes in case of the emitter for linearly polarized radiation \cite{Dreyhaupt2005} and concentric rings for the radial polarization \cite{Winnerl2009}, respectively, is patterned on semiinsulating GaAs. The divergent THz beam is collimated and refocused by a pair of polymethylpentene (TPX) lenses with a focal length of 75 mm. Two 200 $\mu$m thick ZnTe crystals are palced in the focal plane alternatingly for electro-optic detection. A (110) oriented crystal is employed for registering the transverse field components, while a (100) oriented crystal is used for detecting the longitudinal field components \cite{Planken2001}. The orientation of the crystal axis is confirmed by x-ray diffraction and by investigating the THz signal dependence on the azimuthal orientation of the crystals \cite{comment3}. Changing the time delay between the pulse exciting the THz emitter and the probe pulse allows for registering THz transients. By scanning the detection unit, consisting of the ZnTe crystals and the optical components to read out the near-infrared polarization state, beam profiles along the $x$-direction (for $y$ = 0) are measured in steps of 0.1 mm. 

\begin{figure}
\includegraphics[width=8cm]{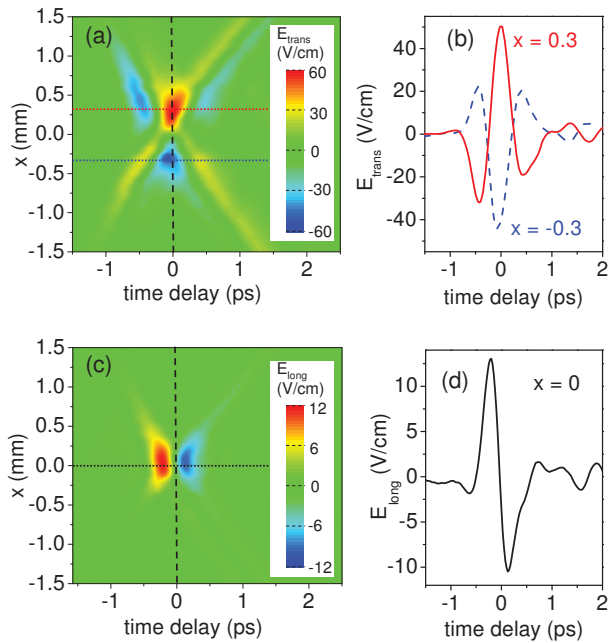}
\caption{\label{fig:epsart} (Color online) Temporal wave fronts for a radially polarized THz pulse. (a) The transverse field components with positive values of the electric field indicating fields pointing in positive x-direction. (c) The longitudinal field components with positive values of the electric field indicating fields pointing in positive $z$-direction. Single THz traces for the $x$-positions indicated by the dotted lines are depicted in panel (b) and (d).}
\end{figure}

In Fig. 2 the measured transverse and longitudinal THz fields are shown for a radially polarized beam. The transverse fields clearly show the feature of a radially polarized single cycle THz pulse: a main THz half cycle with the field pointing radially outwards (i.e. positive THz fields for $x > 0$ and negative THz fields for $x < 0$) around zero time delay, preceded and succeeded by weaker half-cycles with the field pointing radially inwards. The longitudinal field components are strongest around $x$ = 0. For zero time delay, the longitudinal field vanishes for all values of $x$. This  indicates that all frequency components of the single-cycle THz pulse obey the phase shift of $\pi/2$ between longitudinal and transverse fields.

For demonstrating this more quantitatively the frequency dependent amplitude and phase are calculated by Fourier transformation of the THz transients. As one can see in the inset of Fig. 3, the spectral content of the longitudinal and transverse components is similar for a fixed location $x$. The phase difference   $\Delta \Phi =  \Phi_{trans} -  \Phi_{long}$ between the longitudinal and transverse component is found to be $\pi/2$ for all reasonably strong frequency components.

\begin{figure}
\includegraphics[width=6cm]{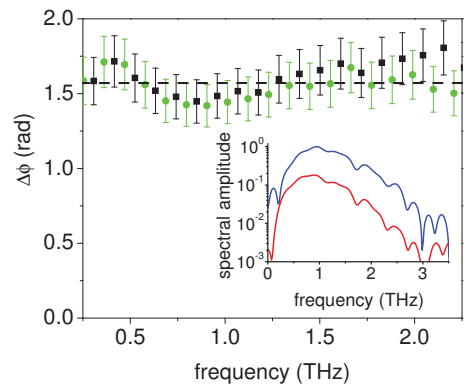}
\caption{\label{fig:epsart} (Color online) Phase difference of longitudinal and transverse fields. The black squares are data measured for $x$ = 0.1 mm, the green dots measured for $x$ = 0.2 mm. The dashed line indicates the value $\pi/2$. In the inset the amplitude spectra of the longitudinal (red) and transverse (blue) components measured at $x$ = 0.2 mm are depicted.}
\end{figure}

\begin{figure}
\includegraphics[width=7cm]{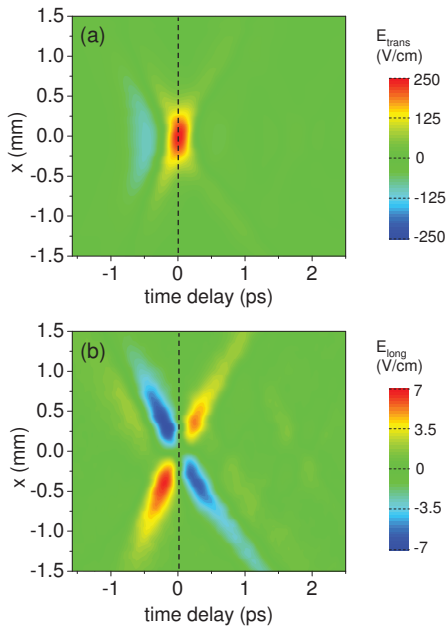}
\caption{\label{fig:epsart} (Color online) Temporal wave fronts for a linearly polarized THz pulse. (a) The transverse field components with positive values of the electric field indicating fields pointing in positive x-direction. (b) The longitudinal field components with positive values of the electric field indicating fields pointing in positive z-direction.}
\end{figure}

We now investigate the much more common mode of linearly polarized THz beams. The transverse components are characterized by the common Gaussian beam with a maximum for $x$ = 0 and $t$ = 0 (Fig. 4a). The measured pattern for the longitudinal field (Fig. 4b) exhibits extrema around $x$ = 0.3 mm, resembling the expected structure (cf. Fig. 1b for $\omega t =  \pi/2$ and $\omega t =  3/2\pi$). Another scan was performed in the $y$-direction, i.e. perpendicular to the direction of polarization. In this case a similar transverse field profile but vanishing longitudinal fields were observed (not shown). Comparing Figs. 4a and 4b we find that the maximum of the transverse field occurs at the time where the longitudinal field is zero for all values of $x$. Hence, also all frequency components of the focused linearly polarized beam are characterized by a phase shift of $\pi/2$ between transverse and longitudinal fields. We note that the X-like pattern seen in the diagrams are signatures of diffraction limited broadband pulses. Such patterns have been observed for single-cycle THz pulses emitted from antennas \cite{Bitzer2008} and are also seen in calculated patterns for radially polarized radiation from accelerator based sources \cite{Tavalla2011}.

\begin{figure}
\includegraphics[width=5cm]{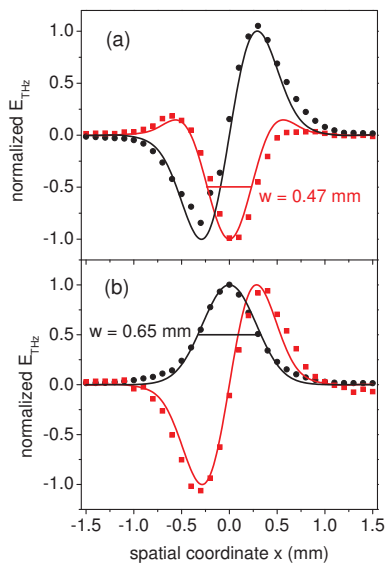}
\caption{\label{fig:epsart} (Color online) Beam profiles for (a) radially and (b) linearly polarized beams. The red squares correspond to the measured longitudinal fields, the black circles to the transverse fields. The lines represent calculated beamprofiles (see main text).}
\end{figure}

Next, the results of the experiments involving radially and linearly polarized beams are compared with each other and with theory. In Fig. 5 the beam profiles for the two field components are depicted. The profiles are selected for the times, where the corresponding component reaches maximum amplitude. The experimental data are compared with calculations where we apply the formalism for focusing vectorial beams beyond the paraxial approximation \cite{Richards1959, Youngworth2000}. We describe the input beams in front of the focusing element by a Gaussian profile in case of the linearly polarized beam and by the product of a Gaussian and a Bessel function of first order and the first kind in case of the radially polarized beam \cite{Winnerl2009}. The calculation is performed for $\lambda = 0.3$ mm. The full width at half maximum of the beams in front of the focusing element is 30 mm for both types of beams. This resembles the situation for the central wavelength of the single cycle pulses studied in the experiment. This calculation reproduces the measured fields very well (cf. Fig. 5). Note, however, that in the experiment the radiation is broadband. In both the measured and calculated beam profiles the full width at half maximum $w$ of the longitudinal field of the radially polarized beam is smaller than the width of transverse component in case of linearly polarized radiation. While this effect is expected, there is an important difference to many experiments in the visible spectral range, where only the total intensity is recorded. In the latter case the tighter focusing of radially polarized beams is only observed for very large numerical apertures, since it requires the longitudinal component to be stronger than the transverse components. In our case of direction-sensitive detection of the fields, however, the narrower beam profile can already  be observed for moderate focusing, where the transverse components still exceed the longitudinal components. 

Finally we discuss some implications of our findings. The knowledge of the phase relation between longitudinal and transverse fields can be applied to detect these components selectively, even when the applied technique is insensitive to the field orientation. An example for such a technique is the Kerr-shutter method \cite{Miyaji2004}. For wavelengths in the visible or UV range, where temporal gating becomes difficult due to the short period of the waves, interferometric techniques may be applied in order to exploit the spatial separation of longitudinal and transverse fields caused by the universal phase shift. This may enhance the spatial resolution for example in nano-lithography. Moreover, our technique can be employed to study fundamental aspects of light-matter interaction. For the intensity-based experiments in the visible range it has been debated controversially, whether a photodetector can register the longitudinal field in the center of a radially polarized mode, since the latter is accompanied by a vanishing magnetic field component and thereby a vanishing Poynting vector \cite{Dorn2003, Miyaji2004}. The possibility to register amplitude and phase of transverse and longitudinal THz field components should allow one to gain insights into this interesting issue, e.g. by probing the transmission through quantum wells. Intersubband transitions in quantum wells require a field component perpendicular to the quantum-well plane. Hence, electrons in quantum wells will interact strongly with the longitudinal field, when the quantum wells are placed on the electro-optic sensor crystal. With accelerator-based sources \cite{Tavalla2011} capable to generate radially polarized fields in the MV/cm range even the high-field regime can be investigated.


In conclusion, by directly recording transverse and longitudinal field components of single cycle THz pulses we have provided insights into the spatial and temporal structure of the fields. The observed phase difference of  $\pi/2$ between longitudinal and transverse fields, which can be concluded from a symmetry argument, is of universal nature, i.e. it does not depend on frequency, the type of mode or focusing conditions. It can be applied to selectively detect longitudinal or transverse field components even when the detection mechanism itself is not sensitive to the field direction.  This may lead to a better understanding of light-matter interaction in case of vanishing Poynting vector.

We thank F. Peter and B. Zimmermann for their contribution to the generation of radially polarized THz waves. Discussions with S. Chatterjee and S. Grafström are gratefully acknowledged.


\begin{thebibliography}{99}
\bibitem{comment1}
Of course the phase, or, equivalently, $t$ = 0, is only defined in a relative way. Here we set $\omega t = 0$ corresponding to the maximum of the transverse field pointing outwards and keep this convention throughout the paper. Equivalently the phase $\omega t = 0$ corresponds to the maximum of the transverse field pointing in positive $x$-direction for the linearly polarized beam.
\bibitem{comment2}
We note that there are also transverse components along the $y$-direction in the focal plane (not shown). For the focusing conditions chosen here they are about four orders of magnitudes weaker as compared the components in $x$-direction. Since both transverse components are in phase, the components in $y$-direction also exhibit the $\pi/2$ phase shift with respect to the longitudinal components.
\end{thebibliography}
\end{document}